\documentclass{icrc29}
\usepackage{graphicx,amssymb,amsmath,times}
\newcommand{\citep}{\cite}
\newcommand{\citet}{\cite}

\begin{document}
%Title of paper
\title[Search for Cross-Correlations of UHECR with BL Lacs]
{Search for Cross-Correlations of Ultra--High-Energy Cosmic Rays
With BL Lacertae Objects}
\author[C. Finley, S. Westerhoff, for the HiRes Collaboration] 
{Chad B. Finley$^a$, Stefan Westerhoff$^a$, for the HiRes Collaboration\\
        (a) Department of Physics, Columbia University, New York, New York,
            USA
}
\presenter{Presenter: C. Finley (finley@physics.columbia.edu), \  
usa-finley-C-abs1-he14-oral}
\maketitle

\begin{abstract}

We present the results of searches for correlation between ultra--high-energy
cosmic rays observed in stereo mode by the High Resolution Fly's Eye (HiRes)
experiment and objects of the BL Lac
subclass of active galaxies.  In particular, we
discuss an excess of events correlating with confirmed BL Lacs in the Veron
10th Catalog.  As described in detail in 
\citet{Abbasi:2005bllac}, the significance level of these correlations
cannot be reliably estimated
due to the {\it a posteriori} nature of the search, and the results must be
tested independently before any claim can be made.  We identify the precise
hypotheses that will be tested with independent data.

\end{abstract}

\section{Introduction}

One of the most striking astrophysical phenomena today is the
existence of cosmic ray particles with energies up to and exceeding
$10^{20}$\,eV.  It is currently unknown where and how these
particles are accelerated to such energies.  Among the potential
sources which have been considered are BL Lacertae objects.
BL Lacs are a subclass of blazars, which are active galaxies in which the jet 
axis happens to point almost directly along the line of sight.
Blazars are established sources of high energy $\gamma$-rays
above 100\,MeV \citep{Hartman:1999fc}, and several BL Lac objects
have been observed at TeV energies with ground-based air Cherenkov
telescopes.  High energy $\gamma$-rays could be
by-products of electromagnetic cascades from energy
losses associated with the acceleration of ultra--high-energy cosmic
rays (UHECR) 
and their propagation in intergalactic space \citep{Berezinskii:1990,
Coppi:1996ze}.

Significant correlations between 
subsets of BL Lac objects and cosmic rays observed by the 
Akeno Giant Air Shower Array (AGASA) and 
Yakutsk experiments have been claimed 
\citep{Tinyakov:2001_bllac,Tinyakov:2001_tracing,Gorbunov:2002_egret}.
However, the claims are controversial
\citep{Evans:2002ry,Stern:2005fh}, and in some cases it has been shown that 
statistically independent data sets do not 
confirm the correlations \citep{Torres:2003ee}.  

The operation of the stereoscopic High Resolution Fly's Eye (HiRes)
air fluorescence detector is providing a large data set of cosmic ray
events with unprecedented angular resolution for the study of small-scale
anisotropy and source correlations.  In this paper, we report on searches
for correlations between BL Lac objects and HiRes stereoscopic events 
observed between 1999 December and 2004 January.
The quality cuts applied to this data sample
are described in detail in \citet{apjl2004,apj2005}.

\section{Maximum Likelihood Method}\label{sec:method}

We apply an {\it unbinned} maximum likelihood method in 
the search for UHECR correlations with point sources.
This approach
uses the probability density function for each individual event rather than 
requiring a fixed bin size.  Two important advantages 
of this method are the ability to accommodate events with different errors, 
and to give weighted sensitivity to angular separations---avoiding
the loss of information that follows from 
choosing an angular separation cut-off.
The method is described in more detail in \citet{Abbasi:2005bllac,apj2005}.  

Briefly, 
the premise involved in the maximum likelihood analysis is that the data sample
of $N$ events consists of $n_s$ source events which came from 
some source position(s)
in the sky, and $N-n_s$ background events.  The probability distribution
of arrival directions ${\bf x}$ for a source event is given by 
$Q_i({\bf x},{\bf s})$, which depends on the source location ${\bf s}$ 
and the $i$th event's angular error function.  The probability distribution
of arrival directions for a background event is given by the detector
exposure to the sky, $R({\bf x})$.

Given a set of $M$ source locations, we define the total source probability
distribution $Q_i^{tot}({\bf x})$ for the $i$th event 
as the sum of the individual source probabilities, each weighted by the
detector's exposure to the $j$th source:
\begin{equation}
Q_{i}^{tot}({\bf x}) = \sum_{j=1}^{M} Q_i({\bf x},{\bf s}_j) R({\bf s}_j)
             /\sum_{k=1}^{M} R({\bf s}_k)~~.
\end{equation}
We use $Q_{i}^{tot}$ (rather than $Q_{i}$) 
to define the partial probability and likelihood functions
(see ~\citet{apj2005} for details).  

The best estimate for the number $n_s$ of events contributed by  
the sources can be determined by finding the value of 
$n_{s}$ that maximizes the likelihood ratio $\mathcal{R}$:
\begin{equation}
{\mathcal R}(n_{s}) = \prod_{i=1}^{N}~
        \left\{\frac{n_{s}}{N}
           \left(\frac{Q_{i}^{tot}({\bf x}_{i})}
                      {R({\bf x}_{i})}
           -1\right)
        +1\right\}
\end{equation}
In practice, we maximize $\ln\mathcal R$.  The maximized value of 
$\ln\mathcal R$ is a measure of the deviation from
the null hypothesis ($n_s=0$).  We estimate the significance
by performing the same likelihood analysis on simulated data sets
and ranking them according to their $\ln\mathcal R$ values.
We will use $\mathcal F$ to 
denote the fraction of simulated, isotropic event sets which yield a
value of $\ln \mathcal R$ greater than or equal to that of the data.
It is worth emphasizing that $n_s$ denotes the {\it excess} number
of events correlating with source positions, above the background expectation.

For the source probability function
$Q_{i}$ we employ a circular Gaussian of width $\sigma_i$ corresponding to the
angular uncertainty of the $i$th event, as estimated by the stereo event
reconstruction.  For the background probability function $R({\bf x})$,
we use estimates based on either time-swapping of the events or a full
detector simulation, depending on the number of events in the sample;
these methods are described in detail in \citet{Abbasi:2005bllac,apjl2004}.

\section{Analysis}
\label{sec:newcorr}

The 271 published HiRes events above $10^{19}$\,eV
were recently analyzed in \citet{Gorbunov:2004_hires}, 
and correlations with a sample of 157 BL Lacs from the Veron 10th
Catalog \citep{Veron:10th} were found.  The sample consisted
of the confirmed BL Lacs classified as ``BL'' in the catalog
with optical magnitude $m<18$.
We verify this analysis by applying the maximum likelihood method
to the same data set and source sample,
and find $\ln\mathcal R = 6.08$ for $n_s = 8.0$; 
the fraction of Monte Carlo sets with higher 
$\ln\mathcal R$ is $\mathcal F = 2 \times 10^{-4}$.

The magnitude cut $m<18$ was previously identified 
as enhancing correlations between BL Lacs and the 
AGASA data \citep{Tinyakov:2001_tracing}.
The current HiRes result does not strictly
confirm the previous correlations, however, because the energy threshold
has been lowered.  
Using the same energy threshold of $4\times 10^{19}$\,eV that was 
used for AGASA,
the HiRes data in fact has a deficit of events correlating with this
BL Lac sample.

The result nevertheless warrants further study.  Because it represents
a new claim based on the current HiRes data set, it can only be confirmed 
with new data.  In this paper, we continue the analysis using the current HiRes
data to explore how variations of the hypothesis affect the result.  
We report on three results suggesting well-defined, well-motivated 
hypotheses which can be tested in the future with
independent HiRes data.

{\bf Event Sample --- Low Energy Events:}
Almost all of the events above $10^{19}$\,eV which contribute to the 
observed correlation have energies between $10^{19}$\,eV and $10^{19.5}$\,eV.
At these energies, it is generally assumed that the Galactic 
magnetic field will deflect a proton primary by many degrees; nuclei will
be deflected even more.  In spite of this, 
the correlations are consistent with the $\sim 0.5^{\circ}$ scale 
of the detector angular resolution.  
This would imply that the correlated primary cosmic rays are neutral.  
Since the chief motivation for restricting the analysis to events
above some energy threshold is 
to minimize the deflections by magnetic fields, 
this motivation is removed if the primaries are neutral, and an analysis of
the entire HiRes stereo data set of 4495 events at all
energies is justified.

Applying the analysis to the entire data set and the same sample of BL Lacs, 
we find
correlations at about the same level of significance as originally found
for events above $10^{19}$\,eV only: the analysis gives
$n_s = 31$, with $\mathcal F = 2 \times 10^{-4}$.
This of course includes the effect of the correlated
events above $10^{19}$\,eV;
for the independent sample of 4224 events below $10^{19}$\,eV, we find
$n_s = 22$, with $\mathcal F = 6\times 10^{-3}$.

{\bf Source Sample --- ``HP'' BL Lacs:}
The sample of BL Lacs discussed above includes only 
confirmed BL Lacs which are classified as ``BL'' in the Veron 10th Catalog.
The rest of the confirmed BL Lacs are classified as ``HP'' 
(high polarization).  It is natural to perform the analysis on these
objects; many in fact are among the most luminous BL Lacs.  
We employ the same cut on optical magnitude $m<18$ to the ``HP'' objects, 
which produces a sample of 47 objects.
The result of the maximum likelihood analysis applied to this 
independent sample of BL Lacs and the HiRes events
above $10^{19}$\,eV is $n_s = 3.0$, with $\mathcal F = 6 \times 10^{-3}$.  
We also perform the same analysis on the events below
$10^{19}$\,eV.  No excess is found.

A summary of the results that are statistically independent is given in 
Table~\ref{tab:hiresbl}.  We have also performed the equivalent analyses 
on the same classes of BL Lacs with $m\ge 18$:
no excess correlation is found in any of these cases.  
It is apparent from these results that the $m < 18$
cut which was identified in \citet{Tinyakov:2001_tracing} 
as optimal for AGASA also isolates the BL Lac objects which show
excess correlations with HiRes events.
Under the BL Lac source hypothesis, of course,
it is not unreasonable to expect the closer and more 
luminous objects to contribute more strongly.  However,
since the Veron catalog is not a uniform sample of BL Lac objects,
the interpretation of this cut may involve a more complicated interplay of
selection effects from the underlying surveys which make up the 
catalog.

{\bf Source Sample --- TeV Blazars:}
Among the closest and brightest of the ``BL'' and ``HP'' 
BL Lacs are six which are confirmed
sources of TeV $\gamma$-rays \citep{Horan:2003wg}. 
Five of these are high in the northern sky and well within the field
of view of HiRes.
We perform the maximum likelihood analysis on this set of objects
using all of the HiRes data, and find $n_s = 5.6$ with $\mathcal F = 10^{-3}$.
For just the HiRes events above $10^{19}$\,eV,
the result is $n_s=2.0$, with $\mathcal F=2\times 10^{-4}$.

\section{Results and Discussion}

Using an unbinned maximum likelihood method,
we have verified the observation in~\citet{Gorbunov:2004_hires}
that the set of HiRes stereo events with energies above
$10^{19}$\,eV shows correlation with confirmed BL Lacs 
marked as ``BL'' in the Veron 10th Catalog.  We emphasize that the 
observed correlation does not confirm a previous claim, because it  
requires a lower energy threshold.  It can only be confirmed with new data.

We have explored the extension of the analysis to 1) HiRes events of all 
energies, and 2) the rest of the confirmed BL Lacs (labeled ``HP'') in
the Veron 10th Catalog.  In each case, correlations at the significance level
of $\sim 0.5\%$ are found.  While statistically independent from the
above result, these are not strictly tests of that claim.
However, in combination with that claim they offer well-defined hypotheses
which can be tested with new data.  The combined results are summarized
in Table~\ref{tab:summary}.  Also shown
are the results for HiRes events and the subset
of BL Lacs which are confirmed sources of TeV $\gamma$-rays.

The HiRes detector will continue observations through the end of 
2006 March.  By that time the independent sample of data since
2004 January is expected to reach 
approximately 70\,\% of the size of the sample analyzed here.  
This will provide an 
opportunity to test the correlations in Table~\ref{tab:summary}.  
We note that while the correlation signals appear stronger for the 
events above $10^{19}$\,eV, 
a conservative approach which includes consideration of the entire 
data set will help to avoid the possibility that a real correlation has been 
``over-tuned'' by an arbitrary threshold and is missed in a future analysis.

\begin{table}
\begin{minipage}[t]{7.5cm}
{\small
\begin{center}
\begin{tabular}{cccccc}
\hline
\hline
BL Lacs       & HiRes En-     & \multicolumn{2}{c}{Results} \\[-2pt]
\cline{3-4}
(\# Obj.)     & ergies [EeV]      & $n_s$ & $\mathcal F$ \\ 
\hline
``BL'' (157)   & $E>10$  & 8.0  & $2\times 10^{-4}$  \\
               & $E<10$  & 22.  & $6\times 10^{-3}$  \\
``HP'' (47)    & $E>10$  & 3.0  & $6\times 10^{-3}$   \\
               & $E<10$  & (0)  & 0.5    \\
\hline
\hline
\end{tabular}
\caption{\label{tab:hiresbl}Correlation results between HiRes events 
above or below 10~EeV and 
confirmed BL Lacs with $m<18$ in the Veron 10th Catalog, classified either as 
``BL'' or ``HP'' (high polarization).  The estimated number of source events
is $n_s$; $\mathcal F$ is the fraction 
of simulated HiRes sets with stronger correlation signal. 
The four cases are non-overlapping.}
\end{center}
}
\end{minipage}
\hfill
\begin{minipage}[t]{7.5cm}
{\small
\begin{center}
\begin{tabular}{ccc}
\hline
\hline
BL Lacs                & \multicolumn{2}{c}{HiRes Sample} \\
(\# Obj.)              & All Energies       & $E>10$\,EeV \\
\hline
``BL'' (157)            & $2\times 10^{-4}$  & $2\times 10^{-4}$ \\
``BL''+``HP'' (204)   & $5\times 10^{-4}$  & $10^{-5}$ \\
TeV Blazars (6)       & $10^{-3}$          & $2\times 10^{-4}$ \\
\hline
\hline
\end{tabular}
\caption{\label{tab:summary}Correlation results between HiRes events
(all energies and above 10 EeV only) and subsets of confirmed BL Lacs 
(with $m<18$).  
Shown in each case is the fraction $\mathcal F$ of simulated HiRes sets
with stronger correlation signal than observed using the current data.
These results will serve as hypotheses to be tested with new data.
The samples overlap and are {\it not} independent.}
\end{center}
}
\end{minipage}
\hfill
\end{table}

\section{Acknowledgments}
This work is supported by US NSF grants PHY-9321949,
PHY-9322298, PHY-9904048, PHY-9974537, PHY-0098826,
PHY-0140688, PHY-0245428, PHY-0305516, PHY-0307098,
and by the DOE grant FG03-92ER40732. We gratefully
acknowledge the contributions from the technical
staffs of our home institutions. The cooperation of
Colonels E.~Fischer and G.~Harter, the US Army, and
the Dugway Proving Ground staff is greatly appreciated.

\end{document}